\documentclass[czech]{ExcelAtFIT}

\hypersetup{
	pdftitle={Paper Title},
	pdfauthor={Author},
	pdfkeywords={Keyword1, Keyword2, Keyword3}
}

\ExcelYear{2025}

\PaperTitle{VoteMate: A Decentralized Application for Scalable Electronic Voting on EVM-Based Blockchain}

\Authors{Ivan Homoliak, Tomáš Švondr}
\affiliation{*%
  \href{mailto:xsvond00@vutbr.cz}{xsvond00@vut.cz},
  \href{mailto:homoliak@fit.vut.cz}{homoliak@fit.vut.cz},
  \textit{Faculty of Information Technology, Brno University of Technology}
  }

\Abstract{
Voting represents a cornerstone of democratic societies, enabling citizens to express their will and make collective decisions. As technology advances, societies follow along with it, and the topic of online voting is quickly gaining traction and becoming a significant trend. 
And for a good reason too, it can eliminate the need to print ballot papers or open polling stations—voters can vote from wherever there is an Internet connection. 
However, despite these benefits, online voting solutions should be viewed through a sceptical lens and approached with a great deal of caution as they in the same package of great benefits carry many new threats. 
 
 A single vulnerability could create an opportunity to manipulate the election on a massive scale\cite{jafar2021blockchain}.
 While the centralized systems can offer a high degree of security and defence against an external adversary, what if the government running the system becomes a threat to the democratic system and begins to manipulate the system to its own advantage? Just like that, the problem of democratic elections becomes one of trust. 

 Electronic voting systems must be legitimate, accurate, safe, and convenient when used for elections. The centralized systems may offer all of these attributes, but they lack in transparency, and offer only limited confidentiality. 
 
 Blockchain-based voting represents an elegant substitute for conventional systems.
 Such systems are transparent thanks to their end-to-end verifiability, tamper resistant due to their immutability. 
 And on top of that, they also offer a high level of security. The only thing left is confidentiality, however, by implementing an additional layer of cryptography to the blockchain, we can achieve even this feature of our system.
}

\begin{document}
\maketitle
\section{Introduction}
In our work we explore the practical applicability of blockchain technology in large-scale electronic elections. 
Our system utilizes the SB-vote protocol\cite{SBVote}, which is an end-to-end verifiable self-tracking voting system built on the Ethereum blockchain. 
The voting protocol offers 1-out-of-$k$ voice of candidates and has a self-recovery ability. 

The system relies on noninteractive zero-knowledge proofs (NIZKs) to ensure that each encrypted vote is well-formed, i.e., it encodes a valid choice from the candidate list, without revealing which one. 
This mechanism preserves voter anonymity while enabling third parties to verify the correctness of every vote without trusting any central authority.

Our solution integrates centralized access control, maintained by the election authority, with a decentralized application that facilitates user interaction with the voting protocol. 
This enables voter authentication and restricts participation to eligible individuals - those who have successfully verified their identity - while preserving full confidentiality and anonymity of each vote.

\subsection{Alternatives}
Our system does not exist in isolation and with the whole space evolving rapidly, numerous research teams and organizations around the world have developed or are actively developing a wide range of electronic voting solutions.

Among the most established solutions are government-run centralized systems, such as Estonia’s i-voting platform. 
Although these systems have achieved notable operational success and public adoption, they inherently carry several limitations stemming from their centralized architecture.
In contrast to decentralized, blockchain-based approaches, their verifiability and transparency are often constrained, as critical components of the process remain hidden from public audit. In addition, they introduce a single point of failure, making them susceptible to technical faults or malicious interference.
Most importantly, they rely heavily on trust in central authority, not only to preserve voter anonymity and ballot confidentiality, but also to ensure that the election process remains free from manipulation.

In response to these shortcomings, many initiatives have turned to decentralization and the blockchain technology and cryptographic protocols. 
Although these systems address the trust and auditability concerns of centralized architectures, they often struggle with scalability and practical deployment in large-scale real-world elections. 

An example of such a system is the BBB voting\cite{BBB} platform, which is built upon cryptography-based protocols like the Open Vote Network\cite{hao2010anonymous} and elegantly integrates blockchain technology to enhance overall robustness, transparency, and resistance to manipulation.
The  A key feature of this protocol is that it stores on-chain only data critical to the verification while performing most of the computationally intensive tasks off-chain, which immensely helps overcoming blockchain's limitations and optimizes the process. 
Despite this feature, the system faces scalability challenges and is generally suitable only for small-scale elections, such as boardroom or organizational votes.
This problem of is partially addressed in the SBvote protocol, which is an extended, more scalable version of the BBB voting and is the underlying protocol of our decentralized voting platform.

Another promising blockchain-based approach comes from a project called Semaphore\cite{semaphore}, a platform for zero-knowledge signaling on the Ethereum blockchain.
While it is not exactly a voting protocol like BBB-voting or SBvote (. It enables users to submit a message, or cast a vote, as a verifiable member of a group without revealing their identity.
The primary bottleneck of the system lies in the computational overhead of generating zk-SNARK proofs for large groups, as well as the inherent throughput limitations of the underlying blockchain.
However, its extensive use of zk-SNARKs also provides significant advantages, such as highly efficient and inexpensive verification, making it a strong candidate for privacy-preserving voting and signalling applications.\cite{semaphore}.

\section{The protocol}
\label{sec:Conclusions}
\textbf{SBvote} is a fully verifiable self-tallying voting protocol for anonymous voting with 1-out-of-$k$ selection options.
It utilizes zero-knowledge proof-based verification mechanisms, which are executed via smart contracts directly on the blockchain. The basic protocol is divided into five phases:
\begin{itemize}
    \item \textbf{Voter Registration} – Voters prove their identity to the election authority and submit the address of their cryptocurrency wallet. Once the registration phase is complete, the authority records these addresses in the smart contract and grants eligible voters access to the subsequent phases.
    \item  Additionally, the authority divides voters into multiple voting groups and deploys a separate booth contract for each group.
    
    \item \textbf{SignIn Phase} – Eligible voters commit to participation by submitting to the booth smart contract assigned to their group.

    \item \textbf{Pre-Voting Phase} – The authority performs key generation for multi-party computations.
    \item \textbf{Voting} – Voters submit their blinded (encrypted) votes.
    \item \textbf{Fault recovery} - An optional recovery phase is triggered if a voter, after committing to participate, fails to submit their vote. In this phase, the system mitigates the issue by allowing the remaining participants to contribute recovery data, effectively excluding the non-responsive voter and enabling the tally to proceed without them.
    \item \textbf{Tallying} – The votes are counted, and the result is verified against the smart contract..
\end{itemize}
\section{The Decentralized Application}
\label{sec:Introduction}
The voting application consists of three primary components:
\begin{itemize}
    \item \textbf{Frontend application} – Codenamed VoteMate, this component handles user interactions and mediates communication with the backend server and the blockchain layer.
   \item \textbf{Backend application} – A centralized coordination server responsible for user authentication and enforcing the voting schedule.
    \item \textbf{Smart contracts} – The core of the voting protocol, deployed on the blockchain to ensure transparency, verifiability, and decentralization.
\end{itemize}

\paragraph{Votemate} is a cross-platform decentralized application built using the Angular framework\cite{angular} and the Web3.js library. 
It serves as the primary user interface, allowing voters to register, cast a vote and view the election outcome. 
The application integrates a built-in cryptocurrency wallet, which enables user seamless interaction with smart contracts in a non-custodial, privacy-preserving manner. VoteMate is designed to operate consistently across desktop and mobile platforms, ensuring accessibility and usability regardless of the user's device.

\paragraph{The backend} is implemented using Node.js\cite{nodejs} and operates as a coordination and scheduling layer.
Although the voting process itself is decentralized, the backend assists in off-chain operations such as identity verification, voter group assignment, and time-based enforcement of protocol phases. 
It also generates the multi-party-computations (MPC) keys, verifies and collects partial tallies from individual booths, and publishes the aggregated result to the central smart contract.

\paragraph{Smart contracts} implement the voting protocol and form the decentralized backbone of the system.
They enable secure elections without requiring a central authority to process or validate individual votes. 
All critical voting operations, including vote submission, verification, and tallying, are handled on-chain, ensuring transparency and auditability of the process.
As a result, the system remains verifiable and resistant to tampering by any single party, even with the limited involvement of a centralized authority.



\section{Conclusions}
\label{sec:Conclusions}
We have developed a system that brings decentralized and privacy-preserving e-voting to users, offering a practical demonstration of how blockchain and zero-knowledge cryptography can be applied to real-world democratic processes.
By leveraging modern distributed technologies, we address key challenges such as ensuring transparency, immutability, and verifiability of votes, while preserving the confidentiality of voter choices.
The system is designed as a reliable tool for secure elections, reducing reliance on central authorities and minimising the risk of fraud or manipulation inherent in traditional voting methods.

While not yet a perfect solution — with scalability remaining a primary challenge to support high candidate participation — the system demonstrates that electronic voting can be implemented with strong security guarantees, minimal risk of manipulation, and without compromising either transparency or voter privacy. 
This shows that blockchain technology can clearly be the right path forward for secure and trustworthy e-voting systems.

\section*{Acknowledgements}
I would like to thank my supervisor doc. Ing. Ivan Homoliak, Ph.D., for his help, support and valuable insights. 

\phantomsection


\end{document}